\documentclass[conference]{IEEEtran}
\IEEEoverridecommandlockouts
\usepackage{cite}
\usepackage{algorithmic}
\usepackage{graphicx}
\usepackage[caption=false]{subfig}
\usepackage{textcomp}
\usepackage{xcolor}
\def\BibTeX{{\rm B\kern-.05em{\sc i\kern-.025em b}\kern-.08em
   T\kern-.1667em\lower.7ex\hbox{E}\kern-.125emX}}

\usepackage{amsmath,amssymb,amsfonts}
\usepackage[linesnumbered,ruled,lined]{algorithm2e}
\usepackage{silence}
\usepackage{booktabs}
\usepackage{pifont}
\usepackage{array}
\newcolumntype{L}{>{\centering\arraybackslash}m{3cm}}

\graphicspath{{/}}

\usepackage{multirow}
\usepackage{arydshln}

\begin{document}

\title{Recurrent Graph Transformer Network for Multiple Fault Localization in Naval Shipboard Systems\\
}







\DeclareRobustCommand*{\IEEEauthorrefmark}[1]{%
  \raisebox{0pt}[0pt][0pt]{\textsuperscript{\footnotesize #1}}%
}

\author{
\IEEEauthorblockN{Quang-Ha Ngo\IEEEauthorrefmark{1}, Isabel Barnola\IEEEauthorrefmark{2}, Tuyen Vu\IEEEauthorrefmark{1}, Jianhua Zhang\IEEEauthorrefmark{1}, Harsha Ravindra\IEEEauthorrefmark{2}, Karl Schoder\IEEEauthorrefmark{2}, Herbert Ginn\IEEEauthorrefmark{3}}
\IEEEauthorblockA{\IEEEauthorrefmark{1}\textit{Clarkson University}, \IEEEauthorrefmark{2}\textit{Florida State University}, \IEEEauthorrefmark{3}\textit{University of South Carolina}} \\
Email: \{hanq, jzhang, tvu\}@clarkson.edu, \{ibarnola, hravindra\}@fsu.edu, schoder@caps.fsu.edu, ginnhl@cec.sc.edu
}

\maketitle

\begin{abstract}

The integration of power electronics building blocks in modern MVDC 12kV Naval ship systems enhances energy management and functionality but also introduces complex fault detection and control challenges. These challenges strain traditional fault diagnostic methods, making it difficult to detect and manage faults across multiple locations while maintaining system stability and performance. This paper proposes a temporal recurrent graph transformer network for fault diagnosis in a Naval MVDC 12kV system. The deep graph neural network employs gated recurrent units to capture temporal features and a multi-head attention mechanism to extract spatial features, enhancing diagnostic accuracy. The approach effectively identifies and evaluates successive multiple faults with high precision. The method is implemented and validated on the MVDC 12kV shipboard system designed by the ESDRC team, incorporating all key components. Results show significant improvements in fault localization accuracy, with a 1-4$\%$ increase in performance metrics compared to other machine learning methods.

\end{abstract}


\section{Introduction}
\subsection{Motivation and problem statement}
The growing power demands of propulsion, ship services, power converters and loads necessitate the development of Medium Voltage DC (MVDC) 12kV shipboard system for the design of future Navy fleets and the assessment of the integrated systems \cite{lindahl2018shipboard}. The shipboard MVDC system, known as a highly integrated DC system, directly connects power converters to several loads with power levels ranging from W to MW \cite{castellan2018review}. This architecture raises additional concerns about the impact of a fault on converters and loads connected to the starboard or port bus, as these devices are sensitive to overcurrent and can be damaged by fault conditions \cite{satpathi2018dc}. Simultaneous or rapidly successive fault currents can cause severe damage, compromise system reliability, and pose significant threats, potentially leading to deteriorated system responses. However, fault detection and location in zonal shipboard systems face several significant challenges. One issue is the difficulty in distinguishing faults at the component level due to the increased number of power converters and loads that need to be managed \cite{babaei2018survey}. Additionally, there is an inadequate understanding of how to manage successive multiple pole-to-pole faults, which can lead to cascading failure risks in Naval shipboard systems. To this end, effective fault management in shipboard MVDC systems becomes crucial to identify faults at an early stage and take appropriate actions to minimize their effects, enabling reliable, safe, and robust restoration.


\subsection{State of the art}
Researchers have proposed numerous studies in recent years
to detect faults on MVDC systems. Typically, these schemes on these studies can be classified into traditional and emerging techniques, proposed in the literature \cite{babaei2018survey, aboelezz2024novel, djibo2021fault, satpathi2017short, james2017intelligent, latorre2024pole, doerry2022medium}. Traditional methods primarily rely on analyzing the physical equations governing current flow, estimating impedance, or utilizing overcurrent relays to identify fault locations. Overcurrent relays activate when current exceeds a threshold and are commonly used in SPS. Although they serve as backup for unsymmetrical AC faults, their complex coordination and tightly-coupled networks make them unsuitable for primary fault detection in ship systems \cite{gong2005integrated}. The current differential protection detects internal faults by monitoring the sum of currents entering a component, offering high precision and sensitivity. However, it requires communication links between terminals, and is prone to transformer and pilot wire issues, limiting its practicality for SPS \cite{babaei2018survey}. Another fault location technique uses active impedance estimation by injecting a short current spike via a power converter to evaluate bus impedance through transient response analysis \cite{6422382}. Yet, accurately measuring system impedance quickly during operation remains challenging. The traveling wave methods locate faults by timing high-frequency waves generated by a sudden voltage drop at the fault site \cite{liang2019fault}. Nonetheless, its accuracy in MVDC systems is constrained by the challenges of precise time measurement and factors such as cable terminations.

These emerging methods for fault detection and location in MVDC systems utilize advanced signal processing techniques or machine learning algorithms to analyze complex patterns and features in electrical data. Maqsood et al.\cite{9042408} propose using short-time Fourier transform to detect faults in naval shipboard power systems by extracting unique time-frequency features. Liu et al. \cite{liu2021fault} introduce a multilevel Iterative LightGBM for binary fault diagnosis in shipboard MVDC systems by adjusting sample weights based on classification confidence. Chanda \cite{chanda2011ann} presents an ANN-based method for locating faults on MVDC shipboard power system cables using transient voltage and current waveforms. Li et al. \cite{li2014fault} are the first to integrate wavelet transform with artificial neural networks to classify fault types in MVDC systems by analyzing current signals across multiple frequency bands. Inherently, Senemmar et al. \cite{senemmar2022non} combine wavelet transforms with convolutional neural networks (CNNs) to improve accuracy in non-intrusive load monitoring for two-zone MVDC shipboard power systems. Additionally, Nourmohamadi et al. \cite{nourmohamadi2021fault} employ an active grid impedance estimator to evaluate impedance and an ANN classifier to identify fault types and locations in a simplified MVDC ship system model. Peng et al. \cite{peng2024fault} propose a ResNet18 model to enhance feature extraction for detecting faults in dual-bus MVAC shipboard systems. These advances in data driven methods allow for more effective fault diagnosis. However, they fall short on fault diagnosis in the shipboard system due to poor generalization, making them inadequate to capture complex patterns and features in the data \cite{10296852}. 


Graph neural networks (GNNs), capable of dealing with the unevenly distributed and dynamical power system topologies, are initially applied in physical fault detection \cite{li2022emerging}. The potentials in learning spatial and temporal data is significant, prompting researchers to shift from conventional models to GNN. The papers \cite{pandey2022graph} and \cite{jacob2021fault} propose fault detection and location methods using graph convolutional networks (GCNs) for low-voltage DC microgrids and an 8-bus shipboard power system, respectively. Liu et al. \cite{liu2020deep} introduce a generative adversarial network for fault detection and localization in shipboard power systems, achieving high accuracy with real-time FPGA implementation. Senemmar et al. \cite{senemmar2024non} present a wavelet graph neural network for fault detection and localization in shipboard systems, demonstrating high accuracy even with pulse loads and noise. Diendorfer et al. \cite{8001032} offer a graph-based traversal approach for automating fault detection, identification, and recovery in MVDC systems, validated through HIL simulations with a focus on adaptability and scalability. Additionally, Nguyen et al. \cite{10119158} introduce recurrent graph neural networks for fault diagnostics using voltage measurements in distribution systems. The findings of these studies show that GNN can outperform other data driven methods in fault detection and classification. However, due to the interdependent nature of the system components, distinguishing faults at the component levels or those occurring in rapid succession in MVDC shipboard systems is still challenging.

The research gap identified in the studies can be summarized as follows: first, there is a need for more precise and efficient methods for detecting and classifying fault locations at the component level, as existing techniques such as wavelet analysis, deep convolutional neural networks, and graph neural networks primarily address zonal-level faults and lack detailed accuracy, sensitivity, and specificity. Second, there is a lack of research on fault detection in scenarios involving simultaneous faults, which impacts real-world situations where multiple faults occur in quick succession. More comprehensive research is needed to assess the performance of these technologies under such conditions. This study addresses these gaps by introducing a fault diagnostic scheme that leverages both spatial and temporal correlations in graph-based time-series data, utilizing a recurrent graph transformer network for the MVDC 12kV shipboard power system.

\vspace{2mm}

The primary contributions are outlined as follows:
\begin{itemize}
  \item This paper proposes a novel temporal recurrent graph transformer network (RGTN) for line-to-line fault detection at the component level in the MVDC system.

  \item The proposed RGTN leverages a multi-head attention mechanism to effectively captures both spatial and temporal correlations in graph-based time-series data using multi-head attention mechanism, enhancing the accuracy and robustness of fault detection.
  
  \item The proposed model identifies and evaluates successive multiple faults by using current measurements.

  \item The simulation of the MVDC 12kV ship system is implemented, including all components, to ensure thorough analysis and validation of the proposed methods.

\end{itemize}

The rest of the paper is organized as follows: Section II provides a detailed description of the proposed fault diagnostic scheme. Section III presents and discusses the results of applying the method to fault management in the MVDC 12kV shipboard system. Finally, Section IV offers the conclusions.

\section{Methodology}

This section introduces the principles and structures of
the gated recurrent unit and the graph transformer network. The
proposed fault diagnostic approach based on the recurrent graph transformer layer is then described in detail.

\subsection{Gated recurrent unit (GRU)}
GRU is a type of recurrent neural network designed to handle time-series data effectively, particularly in tasks like detection and classification. It is simpler and faster to train than Long Short-Term Memory (LSTM) networks, with fewer parameters, yet it effectively captures long-term dependencies in sequential data \cite{zhang2023gated}.

GRU uses two gates: the reset gate $r_{t}$ and the update gate $z_{t}$ to control the flow of information. The update gate $z_{t}$ determines how much of the previous hidden state $h_{t-1}$ should be carried forward to the next time step, described as follow
\begin{subequations}
\begin{align} 
\label{eq:GRU1}
&r_{t}=\sigma (W_{r}\cdot[h_{t-1}, x_{t}] + b_{r})\\
\label{eq:GRU2}
&z_{t}=\sigma (W_{z}\cdot[h_{t-1}, x_{t}]+ b_{z})
\end{align}
\end{subequations}

The candidate hidden state $\tilde{h}_{t}$ is computed using the reset gate $r_{t}$. The final hidden state $h_{t}$ is then derived by interpolating between the previous hidden state $h_{t-1}$ and the candidate hidden state $\tilde{h}_{t}$, controlled by the update gate $z_{t}$.
\begin{subequations}
\begin{align} 
\label{eq:GRU3}
&\tilde{h}_{t}=\phi (W_{h}\cdot[r_{t} \ast h_{t-1}, x_{t}] + b_{h})\\
\label{eq:GRU4}
&h_{t}= (1-z_{t})\ast h_{t-1} + z_{t}\ast \tilde{h}_{t}
\end{align}
\end{subequations}
where \( W_r \), \( W_z \), and \( W_h \) represent the corresponding weight parameters. The vectors \( b_r \), \( b_z \), and \( b_h \) denote the bias parameters. The operation \([ \cdot, \cdot ]\) represents data concatenation, and ($\ast$) denotes the element-wise product. $\sigma$ represents the sigmoid activation function and $\phi$ denotes the tanh function.


\subsection{Graph convolutional network}
Graph Convolutional Network (GCN) is a deep learning model designed to operate on graph-structured data and learn features and relationships between nodes. It effectively identifies complex patterns by leveraging node information through its local message passing mechanism \cite{kipf2016semi}. The node feature is processed and updated using the propagation rule, which can be expressed as
\begin{align} 
\label{eq:GCN1}
&H^{(l)} = \sigma (\tilde{D}^{-\frac{1}{2}} \tilde{A} \tilde{D}^{-\frac{1}{2}} H^{(l-1)} W^{(l-1)})
\end{align}
where $\tilde{A}$ denotes the adjacent matrix with the self-loops to each node, $\tilde{D}$ is the degree matrix from the matrix $\tilde{A}$. $H^{(l)}$ represents the features at layer $l$, $W^{(l)}$ is the layer-specific weight matrix, and $\sigma$ is a non-linear activation function.



\subsection{Graph transformer network}
\begin{figure}[tp]
    \vspace{-2mm}
    \centering
    \includegraphics[height = 3.8cm, width=7.2cm]{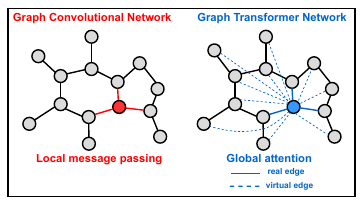}
    \vspace{-2mm}
    \caption{The multi-head attention in a Graph Transformer Network}
    \label{fig:global-attention process}
    \vspace{-1mm}
\vspace{-3mm}
\end{figure}

Graph Transformer Network (GTN), first introduced in \cite{shi2020masked}, can address the limitations of GCN by incorporating a global attention mechanism. As illustrated at Fig. \ref{fig:global-attention process}, the global attention allows GTN to focus on important nodes across the entire graph rather than just local neighborhoods. Specifically, when calculating the attention matrix, both the true edges from the original graph and the virtual edges are included. As a result, GTN is better equipped to handle long-range connections as all nodes are now connected to each other\cite{shi2020masked}. In contrast, GCN relies on local message passing, which aggregates features from multiple neighbors, losing information in the training process. Therefore, GTN is particularly effective in capturing complex relationships and dependencies, making it a more effective model for classification tasks.

To mathematically formalize the process of information aggregation in GTN, the node features at layer $l$, denoted as \( H^{(l)} = \{h^{(l)}_1, h^{(l)}_2, \dots, h^{(l)}_n\} \) are considered. The multi-head attention mechanism for each edge from node \( j \) to node \( i \), including the real and virtual edges, is computed as follows:

\begin{subequations}
\begin{align} 
\label{eq:GTN1}
&q_{c,i}^{(l)} = W_{c,q}^{(l)}.h_{i}^{(l)} + b_{c,q}^{(l)}\\
\label{eq:GTN2}
&k_{c,j}^{(l)} = W_{c,k}^{(l)}.h_{j}^{(l)} + b_{c,k}^{(l)}\\
\label{eq:GTN3}
&e_{c,ij} = W_{c,e}.e_{ij} + b_{c,e}\\
\label{eq:GTN4}
&\alpha _{c,ij}^{(l)} = \frac{\left\langle q_{c,i}^{(l)},k_{c,j}^{(l)}+e_{c,ij}\right\rangle}{\sum_{u\in N(i)}\left\langle q_{c,i}^{(l)},k_{c,u}^{(l)}+e_{c,iu}\right\rangle}
\end{align}
\end{subequations}
where \( \left\langle q,k\right\rangle = \exp\left(\frac{q^\top k}{\sqrt{d}}\right) \) is the exponential scaled dot-product function and \( d \) is the hidden size of each head. For the \( c^{th} \) head attention, the source feature \( h^{(l)}_i \) and distant feature \( h^{(l)}_j \) are transformed into the query vector \( q^{(l)}_{c,i} \in \mathbb{R}^d \) and key vector \( k^{(l)}_{c,j} \in \mathbb{R}^d \), respectively, using different trainable parameters \( W^{(l)}_{c,q} \), \( W^{(l)}_{c,k} \), \( b^{(l)}_{c,q} \), and \( b^{(l)}_{c,k} \). The provided edge features \( e_{ij} \) are encoded and added to the key vector as additional information for each layer.

After obtaining the multi-head attention, the message aggregation is performed as follow:

\begin{subequations}
\begin{align} 
\label{eq:GTN5}
&v_{c,j}^{(l)}=W_{c,v}^{(l)}.h_{j}^{(l)}+b_{c,v}^{(l)}\\
\label{eq:GTN6}
&\hat{h}_{i}^{(l+1)} =\left |  \right |_{c=1}^{C} [\sum_{j\in N(i)}\alpha_{c,ij}^{(l)}(v_{c,j}^{(l)}+e_{c,ij})]
\end{align}
\end{subequations}
where \( \|\cdot\| \) is the concatenation operation for \( C \) head attention. The distant feature \( h_j \) is transformed to \( v_{c,j} \in \mathbb{R}^d \) for the weighted sum.

\subsection{Fault diagnostic scheme using temporal recurrent graph transformer network}

\begin{figure*}[!t]
    \vspace{-4mm}
    \centering
    \includegraphics[height = 9.5cm, width=16.8cm]{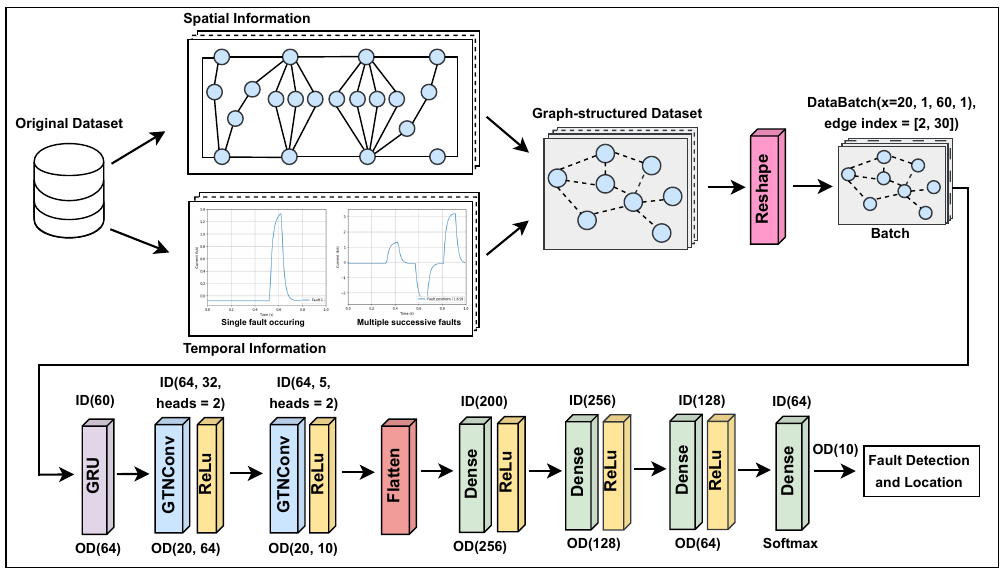}
    \caption{ The Architecture of Recurrent Graph Transformer Network (RGTN) for Fault Localization in the MVDC shipboard system}
    \label{fig:Training_process}
    \vspace{-4mm}
\end{figure*} 

The proposed recurrent graph transformer network model is specifically designed to detect and locate faults in a Naval MVDC 12kV shipboard system at the component level, such as the PGM in zone 3 or the PCM in zone 1. Moreover, it is capable of identifying multiple fault scenarios. The input for the training process consists of current measurements collected at 20 different locations within the system. The currents are denoted as time-series node features $I_{k}$, where $k$ is the length of the period and $k=1, 2, ..., K $. The outputs are categorized into 10 labels, ranging from 0 to 9. Here, 0 represents normal operation without faults, values from 1 to 7 correspond to single faults at one of seven positions, 8 indicates two faults, and 9 signifies three faults occurring in rapid succession. The training process, illustrated in Fig. \ref{fig:Training_process}, can be summarized with the following steps:

\begin{enumerate}
  \item Input pre-processing and graph construction: Current measurements collected from the power system are pre-processed and transformed into graph datasets. Each graph dataset, consisting of 20 nodes and 30 edges, is encoded into input data including the node features, the edge indices, the adjacency matrix. These datasets are then divided into batches and propagated through the GRU layer to capture temporal dependencies.
  \item Model architecture: The proposed model architecture is built with multiple layers. The first layer, a GRU, is used to extract temporal features, followed by two graph transformer layers with the multi-head attention mechanism that facilitates the information propagation and aggregation. Through updating node features iteratively and leveraging global attention mechanisms, GTN can capture complex relationships and learns meaningful representations from graph-structured data.
  \item Training procedure and ouput layer: To compute the probability distribution over fault locations, we employ fully connected layers followed by a SoftMax activation function for accurate fault classification. The proposed model is trained using the cross-entropy loss function, which enhances its ability to detect and locate faults, as well as identify scenarios with multiple faults. By leveraging the SoftMax function, the network can effectively distinguish between normal operation, single faults, and multiple fault occurrences. The loss function $\mathcal{L}(\hat{y}, y)$ can be expressed as

\begin{align}
\label{eq:loss} 
&\mathcal{L}(\boldsymbol{\hat{y}}, \boldsymbol{y}) = -\sum_{i=1}^{n} y_{i} \log \left( \frac{e^{\hat{y}_{i}}}{\sum_{j=1}^{k} e^{\hat{y}_{j}}} \right)
\end{align}
where {\boldmath$y$} is the ground-truth label and {\boldmath$\hat{y}$} is the output label. $y_{i}$ represents the true probability of class $i$ and $\hat{y_{i}}$ represents the predicted probability of class $i$.
\end{enumerate}

By combining GRU with GTNs, the proposed model effectively leverages both temporal features and the graph structure of the MVDC shipboard power system, resulting in accurate fault detection and localization.

\subsection{Performance evaluation}
To assess how the proposed model performs in diagnosing faults within the MVDC 12kV system, we use the following evaluation metrics:

\begin{enumerate}
    \item \textbf{Accuracy:} represents the percentage of correctly classified samples in the test dataset. It is determined by dividing the count of accurate predictions by the overall number of predictions. A higher accuracy score signifies that the model performs well in accurately identifying fault locations.
    \begin{equation}
    \text{Accuracy} = \frac{\text{Number of correct predictions}}{\text{Total number of predictions}}
    \end{equation}

    \item \textbf{Recall:} measures the proportion of actual positive cases  that the model correctly identifies. It is calculated by dividing the number of true positives by the sum of true positives and false negatives.
    \begin{equation}
    \text{Recall} = \frac{\text{True Positives}}{\text{True Positives} + \text{False Negatives}}
    \end{equation}

    \item \textbf{Precision:} measures the ratio of true positive predictions to the total number of positive predictions made by the model. It is calculated by dividing the number of true positives by the sum of true positives and false positives.
    \begin{equation}
    \text{Precision} = \frac{\text{True Positives}}{\text{True Positives} + \text{False Positives}}
    \end{equation}

    \item \textbf{F1-score:} is the harmonic mean of precision and recall. A higher F1-score indicates that the model performs well in both correctly identifying positive cases and minimizing false positives.
    \begin{equation}
    \text{F1} = 2  \times \frac{\text{Precision}  \times \text{Recall}}{\text{Precision} + \text{Recall}}
    \end{equation}

\end{enumerate}

These metrics offer an assessment of the proposed model's effectiveness in detecting and identifying different faults within the MVDC shipboard system, offering a comparative analysis compared to other deep learning approaches.

\section{Results and Discussion} 
\addtolength{\tabcolsep}{11pt}
\setlength{\tabcolsep}{0.4\tabcolsep}
\begin{table}[bp]
\centering
\caption{shipboard power system model overview \cite{team2017model}}
\begin{tabular}{llllll}
\toprule
\small{Parameter}  & & & &\small{Value} \\
\midrule
Distribution voltage & & & & 12  kV \\
Shipboard power generation & & & & 100 MW \\
Propulsion & & & & 72  MW \\
Special Loads & & & & 29.4  MW \\
\bottomrule
\end{tabular}
\label{table:table0}
\end{table}

\renewcommand{\arraystretch}{1.2} 
\begin{table}[bp]
\centering
\caption{Zonal MVDC 12kV System Load Summary (MW) \cite{team2017model}}
\begin{tabular}{|l|c|c|c|c|}
\hline
                & \textbf{Zone 1} & \textbf{Zone 2} & \textbf{Zone 3} & \textbf{Zone 4} \\ \hline
\textbf{PGM}    & -               & 2-MPGM              & 1-MPGM,              & 1-APGM              \\ 
    &                &               & 1-APGM              &                \\ \hline
\textbf{PCM-1A}    & 10.64                & 10.64                 & 9.17                 & 9.17                 \\ \hline
\textbf{PMM}    & -               &  36               & 36              & -               \\ \hline
\textbf{IPNC}    & 2.77               &  3.13               & 3.95              & 1.99               \\ \hline

\textbf{HRRL}   & -               & 20              & -               & -              \\ \hline
\textbf{Mission Loads}   & 2            & 2            & 1.5             & 1.5             \\ \hline

\textbf{Hotel Load}     & 1.47            & 1.65            & 1.65             & 1.54            \\ \hline
\textbf{Cooling Load}    & 1.26            & 2.52            & 1.26             & 1.26            \\ \hline

\end{tabular}
\label{table:table1}
\end{table}

\subsection{MVDC 12kV shipboard model}
\begin{figure}[tp]
    \vspace{-1mm}
    \centering
    \includegraphics[height = 5.1cm, width=8.1cm]{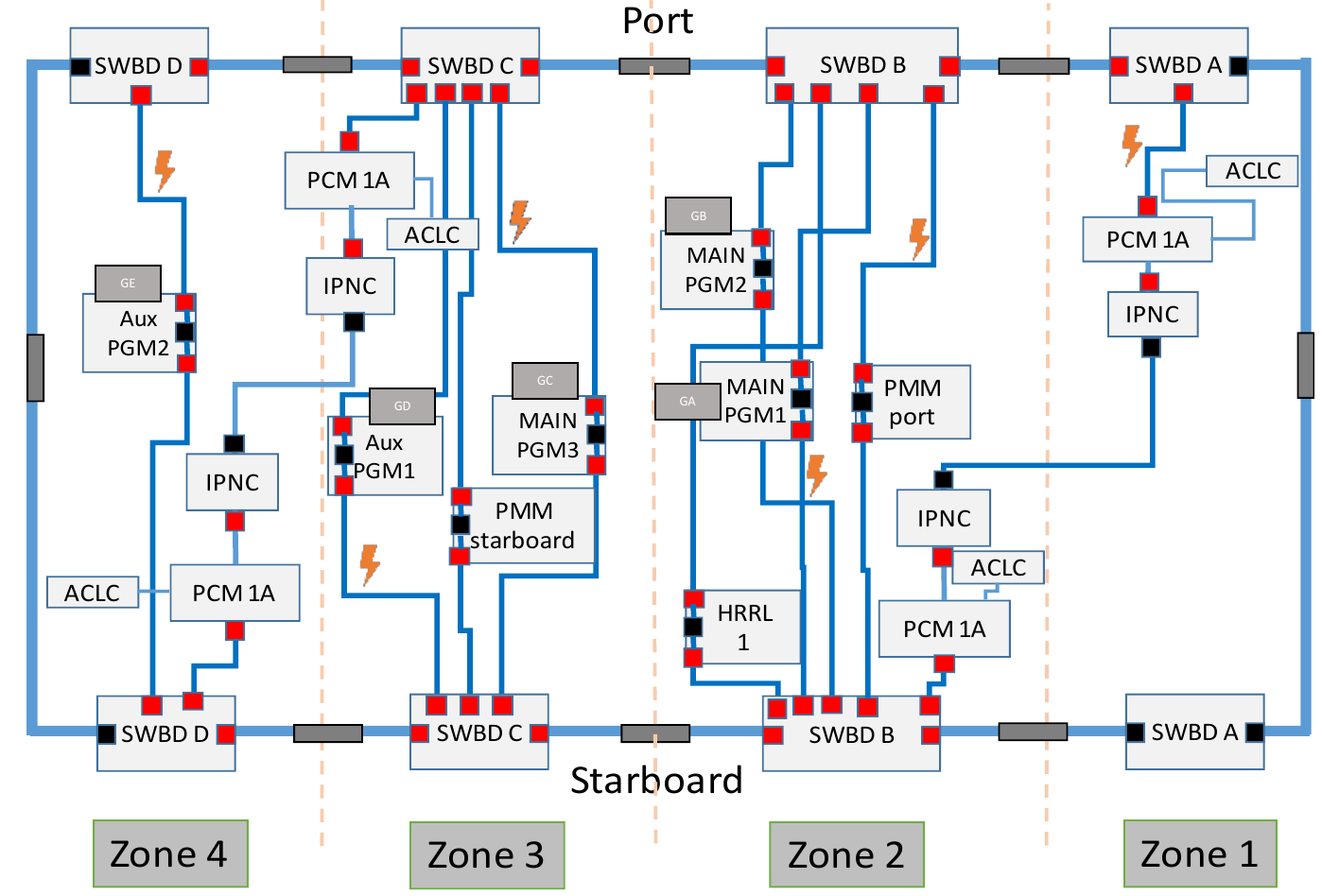}
    \vspace{-2mm}
    \caption{The 4-zonal MVDC 12kV system \cite{team2017model}}
    \label{fig:MVDC 12kV}
    \vspace{-3mm}
\end{figure}

The MVDC 12kV shipboard system is designed to serve as the primary power distribution network in a notional four-zone system architecture. This system is built around a Shipboard Power System (SPS) with a 100 MW rating \cite{vargas2011esrdc}. As shown in Fig. \ref{fig:MVDC 12kV}, each zone of the system includes various modules such as the Power Generation Module (PGM), Power Conversion Module (PCM-1A), Integrated Power Node Center (IPNC), Propulsion Motor Module (PMM), and Energy Storage Module (ESM). Special high ramp rate loads (HRRL) are also incorporated to address specific power demands. Tables \ref{table:table0} and \ref{table:table1}  provide the system overview and breakdown of modules by zones in the SPS model. The SPS model will consist of 3 main PGMs (rated to 30 MW each) and 2 auxiliary PGM (rated to 4 MW each).  One PCM-1A will be modeled in each zone. Mission loads will be modeled separately from the aggregated zonal loads. Zonal loads will be further classified into hotel and cooling loads. The MVDC system operates at a distribution voltage of 12 kV and supports both isolated and parallel bus configurations, enhancing its flexibility and reliability. The power generation is managed by dual-output feed PGMs, which can power port and starboard buses independently, with generators running at higher frequencies (120/240 Hz) and utilizing power electronic converters to limit fault current. Energy storage modules ensure uninterrupted power to critical loads and provide support for special load applications \cite{vu2024real}. The PCM-1A distributes 12 kV MVDC power to various loads at appropriate voltage levels, including 1 kV DC and 450 V AC. The system's modular design, with redundant power feeds and categorized zonal loads, facilitates increased reliability and serviceability, making it a robust solution for modern naval applications.

Multiple faults are more likely to occur in the MVDC 12kV shipboard system due to the complexity and interconnectedness of its components. The system operates with high voltages and integrates various modules such as PGMs, PCM-1As, IPNCs, PMMs, and ESMs, all working together to ensure reliable power distribution. The use of advanced power electronic converters, high-frequency generators, and multiple zones increases the potential points of failure. Additionally, the system's ability to operate in different configurations, such as isolated or parallel bus modes, adds complexity, making it more susceptible to issues like overvoltage, current surges, or component malfunctions \cite{8847803}. The high energy density and the presence of critical loads, including high ramp rate loads and mission-critical systems, further contribute to the risk of multiple faults, as any imbalance or disturbance in one part of the system can quickly propagate and affect other components. This interconnected nature, combined with the high power levels involved, creates a challenging environment for fault management and increases the likelihood of multiple faults occurring rapidly.

\subsection{Fault modelling}
\addtolength{\tabcolsep}{-4.5pt}
\renewcommand{\arraystretch}{1.25}
\begin{table}[bp] 
\vspace{-2mm}
\centering 
    \caption{Dataset Information for Training Procedure}
    \addtolength{\tabcolsep}{-0.2pt}
\begin{tabular}{ *5c } 
\hline
\hline
\textbf{ Elements} & \textbf{ Value} & \textbf{Number} \\
\hline
\multirow{3}{1 em}

\multirow{3}{1 em} \textbf{Fault position\:\:\:\:\:\:\:\:\:\:} & \:\:\:\: 1, 2, 3, 4& 7 \\ 
& \:\:\:\:\: 5, 6, 7 &  \\ 
Multiple faults\: & \:\:(1, 4); (1, 3, 5) &  2 \\ 
Non-fault cases & \:\:\:\:\:\: normal condition & 1  \\
Fault type\:\:\:\:\:\:\: & \:\:\:\:\:line-to-line &   \\
Load scenarios & \:\:\:\:\:\: randomly &   \\
\hline
\multicolumn{3}{c}{\textbf{Total fault cases}: 3,350  $|$ \textbf{Train set}: 2,700 $|$
\textbf{Test set}: 650} & \\
\hline
\hline
\end{tabular}
\label{table:dataset_info}
\end{table}

\begin{figure*}[!t]
    \vspace{-4mm}
    \centering
    \includegraphics[height = 6.9cm, width=17.3cm]{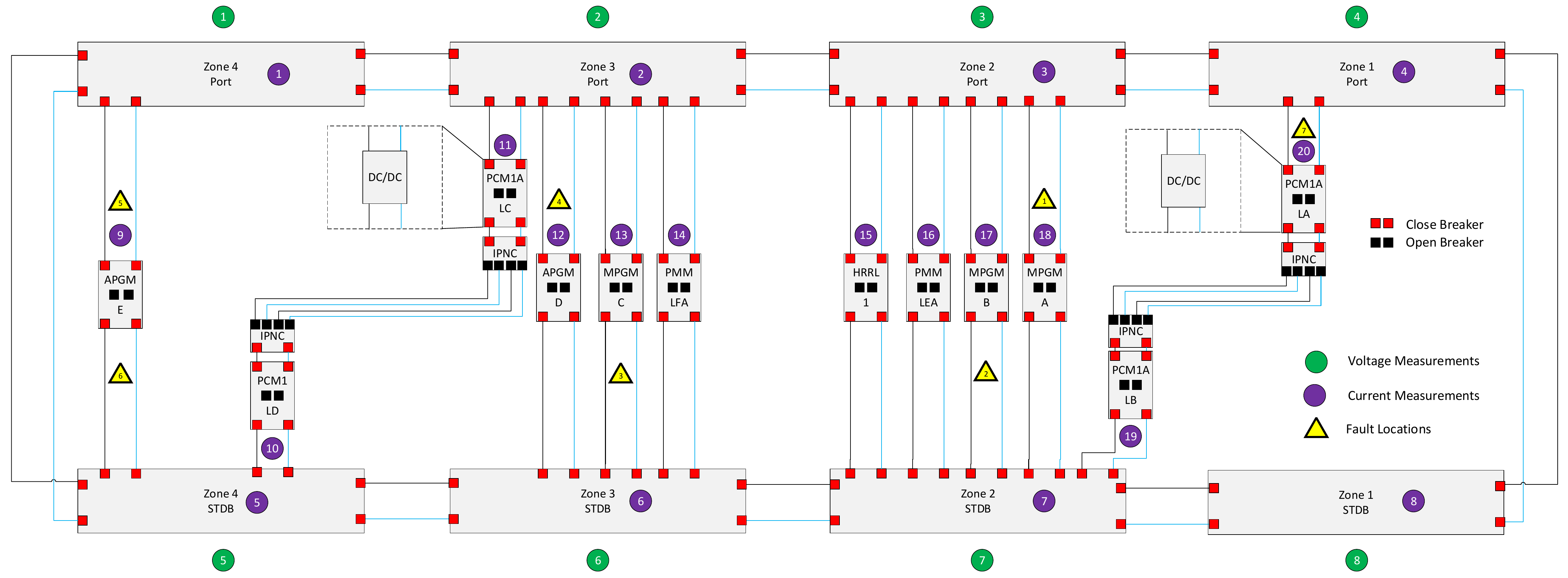}
    \vspace{-3mm}
    \caption{The ESRDC MVDC 12kV System Diagram with Measurement Locations and Fault Scenarios}
    \label{fig:MVDC_simulation}
    \vspace{-4mm}
\end{figure*}

In modeling the MVDC shipboard power system, we focus on implementing the line-to-line short circuit fault due to its prevalence and significance as a type of fault. Locating the faults presents challenges, given the similarities between short circuit faults, particularly because of interconnects among buses and the low impedance of the line. It is also essential to differentiate between single short circuit faults and multiple successive faults, as the latter can occur either simultaneously or in rapid succession, deteriorating system responses. In our model, multiple successive fault scenarios include two or three faults occurring, with each fault happening 50 milliseconds after the previous one. These faults result in a sudden and significant surge in current, accompanied by a decrease in voltage, which our modeling approach effectively captures. Notably, we do not implement line-to-ground faults in the MVDC 12kV system, as we assume there is no grounding in our system. Given the importance of line-to-line short circuit faults, our modeling emphasizes capturing their dynamic behavior and assessing their impact on system performance. 

The line-to-line fault occurs between two lines $L_{1}$ and $L_{2}$, and the fault current can be expressed as
\begin{equation}
I_{f,t} = \frac{V_{L_{1},t}-V_{L_{2},t}}{Z_{f,t} + Z_{L_{1}, t} + Z_{L_{2}, t}}
\label{eq:fault_equation}
\end{equation}
where $V_{L_{1},t}$ and $V_{L_{2},t}$ are the voltages on the lines $L_{1}$ and $L_{2}$ at time \( t \). $Z_{f,t}$ is the impedance of the fault path at time \( t \). $Z_{L_{1}, t}$ and $Z_{L_{2}, t}$ are the impedances of the lines $L_{1}$ and $L_{2}$ at time \( t \), respectively.

\subsection{Simulation and training parameters}
The data used in the proposed diagnostic scheme comes from the dynamic model of the MVDC 12kV ship system, as detailed in Fig. \ref{fig:MVDC_simulation}. Raw data includes current measurements from 20 locations across the system, captured under normal and fault conditions. The load values are randomly adjusted within a range of 70\% to 130\% of their default values specified in the load profile. The ship operates in a four-zone configuration, with each zone containing loads. The MVDC 12kV system is simulated based on the defined parameters and components outlined in the document \cite{team2017model}.
\addtolength{\tabcolsep}{11pt}
\setlength{\tabcolsep}{0.4\tabcolsep}
\begin{table}[bp]
\centering
\caption{Hyperparameter Configuration}
\begin{tabular}{cccccc}
\toprule
\small{Hyperparameters}  & & & & &\small{Value} \\
\midrule
Batch Size & & & & & 8 \\
Epochs for Localization & & & & & 200 \\
Hidden Layer Activation & & & & & ReLU \\
Output Layer Activation & & & & &  Softmax \\
Optimizer & & & & & SGD \\
Initial Learning Rate ($\alpha$) & & & & & 0.01 \\
Dropout Rate  & & & & & 0.1 \\

\bottomrule
\end{tabular}
\label{table:result5}
\end{table}

Fault scenario simulations and data generation for the MVDC 12kV system are conducted using RSCAD. To train the proposed scheme, a dataset is created from these simulations. In each zone, a line-to-line fault simulation lasting 1 second is executed. Current measurements are sampled from all generators, converters, and switchboards at a sampling frequency of 100 Hz. To effectively train the machine learning models, we capture 60 data points over a 0.6-second of 1-second interval, as the system operates under normal conditions for the remainder of the time.

The dataset information of training prcess is summarized in Table \ref{table:dataset_info}. A total of 3,350 samples are collected, with 200 samples of normal operating conditions, 7 fault positions with 350 samples per single fault condition, 350 samples representing two simultaneous faults, and 350 samples representing three simultaneous faults.

Simulations and model training are conducted on a laptop featuring a 2.9 GHz Core i7 processor and 16 GB of RAM. Machine learning models are built using PyTorch library, and the system simulations are implemented in RSCAD. The hyperparameters are summarized in Table \ref{table:result5}. The proposed model is trained for 200 epochs with a batch size of 8. The stochastic gradient descent optimizer, set with a learning rate of 0.01, is used to fine-tune the trainable weights across the different layers of the RGTN.

\addtolength{\tabcolsep}{8pt}
\setlength{\tabcolsep}{0.6\tabcolsep}
\begin{table}[bp]
\centering
\caption{Fault location metrics on MVDC shipboard system}
\begin{tabular}{l cccccc}
\toprule
\small{Method}     & \small{Accuracy} & \small{Recall}  & \small{F1-score} & \small{Precision}\\
\midrule
\small{MLP}  & \small{89.85} & \small{87.93} & \small{84.39} & \small{81.87} \\

\small{GAT}   & \small{90.92} & \small{90.00} & \small{85.49} & \small{83.79}\\

\small{GCN}  & \small{92.46} & \small{92.19} & \small{92.17} & \small{92.18}\\

\small{GTN}   & \small{98.61} & \small{98.45} & \small{98.49} & \small{98.55}\\ 

\small{\textbf{R-GTN}}    & \small{\textbf{99.69}} & \small{\textbf{99.59}} & \small{\textbf{99.64}} & \small{\textbf{99.70}}\\
\bottomrule
\end{tabular}
\label{table:result1}
\end{table}

\setlength{\tabcolsep}{0.6\tabcolsep}
\begin{table}[bp]
\centering
\caption{   Noise Scenario Performance Comparison in MVDC shipboard}
\begin{tabular}{l cccccc}
\toprule
&   \small{  MLP} &   \small{  GAT} & \small{  GCN} & \small{  GTN} & \small{  \textbf{R-GTN}} &\\
\midrule
\small{   No noises} & \small{  89.85} & \small{  90.92} & \small{  92.46} & \small{ 98.61} & \small{  \textbf{99.69}}  & \\
\small{   5$\%$} &  \small{  88.76} & \small{  90.46}& \small{  90.46} & \small{  98.31} & \small{  \textbf{99.23}}  &\\
\small{   8$\%$}  & \small{  87.84} & \small{  88.00} & \small{  89.54}  & \small{  96.15} & \small{  \textbf{97.85}}  &\\
\small{ 10$\%$}  & \small{  84.46} & \small{  83.69} & \small{  86.62} & \small{  94.00} & \small{  \textbf{95.08}} \\
\bottomrule
\end{tabular}
\label{table:result2}
\end{table}






\subsection{Performance of the proposed fault diagnostic model}
To assess the effectiveness of the proposed RGTN model in locating faults within the MVDC 12kV system, two case studies are conducted. These case studies examine different scenarios to assess the model's effectiveness and robustness in addressing various fault conditions.

In the case 1, the study focuses on fault diagnostics with availabel 20 current signals without any noises. By analyzing noise-free data, the study aims to determine the maximum potential of the RGTN model in identifying and locating faults. Regarding the case 2, the study investigates the impact of noises to input measurements on the model's fault detection capabilities. Since practical systems usually experience noise in sensor data and signal transmission, this scenario tests the robustness of the model under such conditions. The goal is to evaluate how effectively the model maintains accurate fault detection performance when confronted with noisy data, offering a more realistic assessment of its practical application.

\subsubsection{No noise}
Table \ref{table:result1} presents a comparative analysis of five models—MLP, GAT, GCN, GTN, and R-GTN—across four performance metrics: accuracy, recall, F1-score, and precision. The proposed model achieves the highest accuracy (99.69$\%$), recall (99.59$\%$), F1-score (99.64$\%$), and precision (99.70$\%$). Notably, R-GTN outperforms traditional models like MLP and GCN, which achieve lower accuracies of 89.85$\%$ and 92.46$\%$, respectively. The superior performance of R-GTN is attributed to the multi-head attention mechanism, which allows the model to better capture global dependencies. In contrast, MLP lacks this mechanism, relying solely on linear layers, while GCN uses the standard neighborhood message passing, limiting its ability to capture complex relationships. Additionally, GAT, which achieves only 90.92$\%$ accuracy, performs worse due to the lack of edge attributes in the input. GTN performs well with 98.61$\%$ accuracy, but R-GTN further improves on these results by incorporating recurrent units and transformer-based graph structures. 

\begin{figure}[t!]
    \vspace{-1mm}
    \centering
    \includegraphics[height = 5.7cm, width=6.5cm]{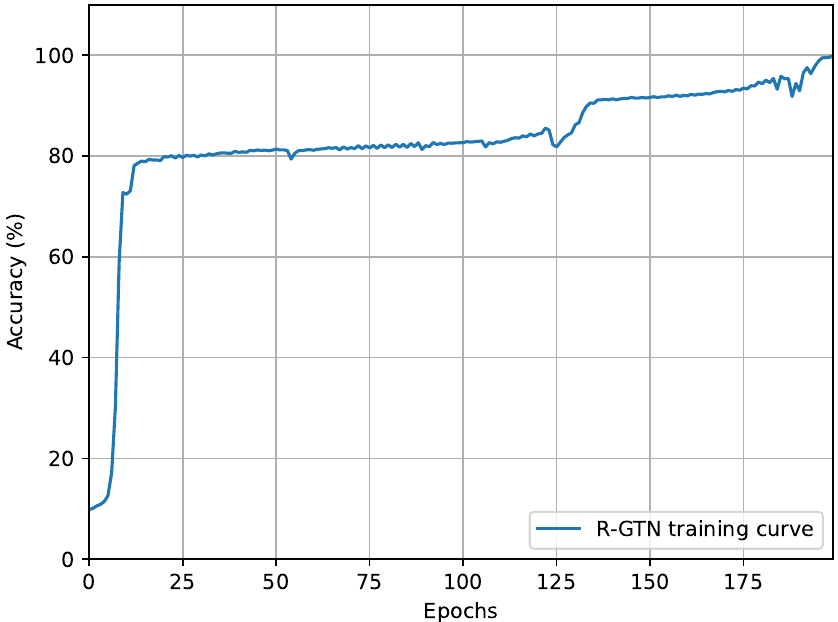}
    \vspace{-3mm}
    \caption{The R-GTN training accuracy for MVDC fault location}
    \label{fig:training curve1}
\end{figure}

\begin{figure}[tp]
    \vspace{-1mm}
    \centering
    \includegraphics[height = 5.7cm, width=6.5cm]{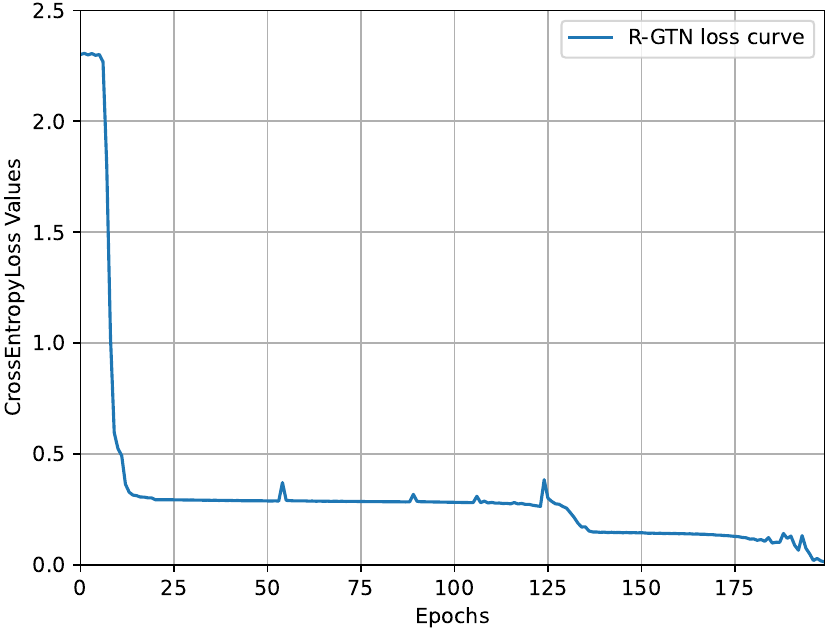}
    \vspace{-3mm}
    \caption{The R-GTN loss curve for MVDC fault location}
    \label{fig:training curve2}
\end{figure}

Fig. \ref{fig:training curve1} visually presents the training curve of the proposed RGTN model in the case 1, highlighting the model's rapid learning from the available signals across all four zones. The curve demonstrates how effectively the model extracts essential information from the time-series current signals, resulting in accurate fault detection. Additionally, the loss curve,  shown in Fig. \ref{fig:training curve2}, demonstrates the model's capability in identifying and localizing fault scenarios throughout the training process.

\begin{figure}[htp]
    \vspace{-1mm}
    \centering
    \includegraphics[height = 5.7cm, width=6.5cm]{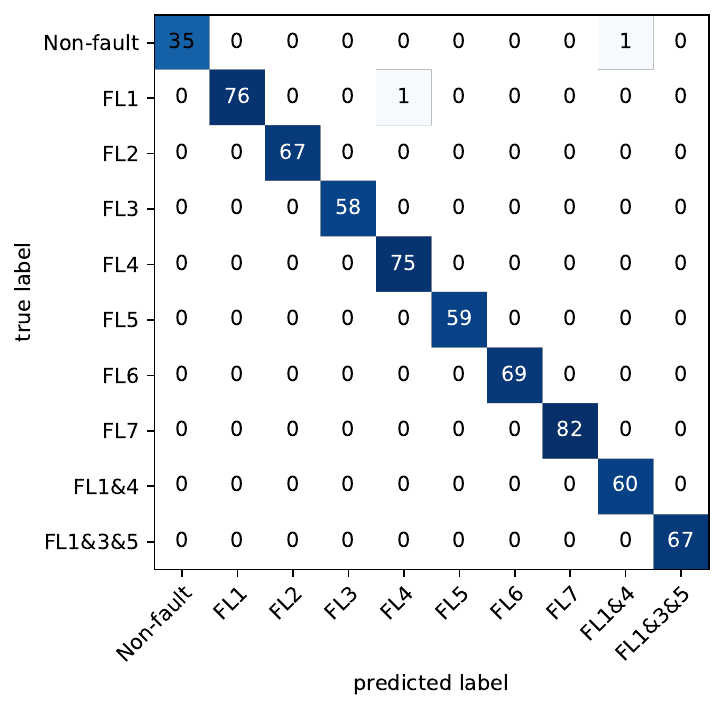}
    \vspace{-3mm}
    \caption{The confusion matrix for fault location in the normal condition}
    \label{fig:training curve3}
\end{figure}

The confusion matrix of RGTN, shown as Fig. \ref{fig:training curve3}, shows strong performance in fault detection across various scenarios. It indicates high accuracy, with most predictions aligning with true labels. Non-fault instances had 35 correct and 1 incorrect prediction, while fault scenarios like FL1 and FL2 also showed high accuracy with  few errors. Other single fault occurring labels such as FL3, FL4, FL5, and FL7 or multiple successive faults have no incorrect predictions. Overall, the matrix underscores the model's reliability in detecting and localizing faults.

The outstanding accuracy and robust fault localization observed in the case 1 highlight the effectiveness of the proposed RGTN model. The following sections will explore and evaluate the model's performance under noise-measurement conditions to evaluate its efficacy for fault diagnosis in the MVDC 12kV shipboard system.

\subsubsection{Impact of measurement noises}
Table \ref{table:result2} presents the accuracy of five machine learning models under three noise-measurement scenarios. R-GTN consistently outperforms the other models across all noise levels. As noise is introduced and increased from 5$\%$ to 10$\%$, all models show a decline in accuracy. However, R-GTN has the best performance, with accuracies of 99.23$\%$, 97.85$\%$, and 95.08$\%$ at 5$\%$, 8$\%$, and 10$\%$ noise levels of input measurements respectively. 

\begin{figure}[htp]
    \vspace{-1mm}
    \centering
    \includegraphics[height = 5.7cm, width=6.5cm]{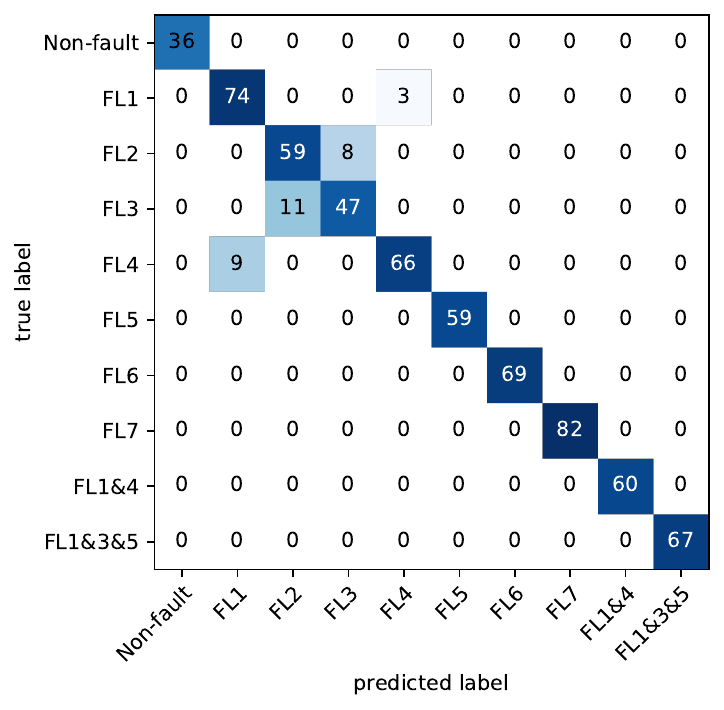}
    \vspace{-3mm}
    \caption{The confusion matrix for fault location with noises}
    \label{fig:training curve4}
\end{figure}

Furthermore, Fig. \ref{fig:training curve4} displays the confusion matrix for this scenario. The model shows strong performance overall, with high numbers on the diagonal indicating correct classifications However, 
There are some misclassifications, notably between FL2 and FL3, where 8 FL2 cases were misclassified as FL3, and 11 FL3 cases were misclassified as FL2. The reason why they mis-classify because location 2 and location 3 connected together cross a switch board. So, when a fault occurs in 1 of 2 location, the characteristics of some current measurements are similar. The model also shows perfect classification for several fault types (FL5, FL6, FL7, two faults occurring, three faults occurring labels) with no misclassifications. Overall, these results of the confusion matrix highlight the robustness and noise resilience of the proposed method for fault localization using currents measurements.

While there was a slight decrease in accuracy compared to previous studies, the approach still yielded valuable insights and successfully identified anomalies. The focus on current measurements, which are fundamental to the system's operation, proved particularly advantageous when dealing with noisy data. These findings indicate that the RGTN model exhibits noise resilience and can effectively localize faults even in challenging and noisy environments.

\subsection{Discussion, limitations, and future works}

   \textit{Comparison with other schemes}: Table \ref{table:result4} presents a comparison of different fault detection methods applied to shipboard systems, showcasing their accuracy and case study focus. Several approaches achieved high accuracy in fault diagnosis. For instance, Multilevel Iterative LightGBM \cite{liu2021fault} demonstrates a 99.77\% accuracy in binary fault diagnosis for MVDC shipboard systems. Wavelet CNN \cite{senemmar2022non} and DNN \cite{li2014fault} methods are also notable, focusing on non-intrusive load monitoring and fault classification with accuracies ranging between 91\% and 97\%. Methods like LSTM \cite{senemmar2021deep} and ResNet18 \cite{peng2024fault} show over 94\% accuracy, emphasizing fault diagnosis in ship systems, particularly for 8-bus and MVAC systems. While these methods achieve significant accuracy, most focus on fault detection at a zonal level for shipboard systems, without addressing more complex scenarios involving detection at the component levels or multiple, successive faults. Our proposed approach achieves 99.69\% accuracy, specializing in multiple fault localization within MVDC 12kV ship systems. This approach not only matches or exceeds the performance of existing methods but also addresses a broader range of fault scenarios, making it more suitable for system management in shipboard environments.
   
   \textit{Limitations and future works}: Although our results have demonstrated the effectiveness of the proposed method, several limitations remain to be addressed. (1) The performance of the method under scenarios with limited observability of current measurements has not been thoroughly evaluated, which is critical for real-world applications. (2) The weighting hyper-parameters of the training process should be optimized to improve accuracy and other metrics, particularly in the presence of measurement noise. (3) Future work should involve Hardware-In-The-Loop (HIL) development to integrate cyber and physical systems for simulating relevant faults and attacks through penetration tests. This will allow for validation of the proposed fault detection methods using a HIL testbed. The testbed will help represent key features of naval systems and provide advanced cyber-physical security features for vulnerability analysis and verification of the proposed algorithms, enhancing the resilience of future naval ship systems.
   
   \textit{Naval Energy Resilience}: The fault detection and localization approach presented in this paper marks a significant advancement in naval shipboard power management. This work not only offers critical improvements but also lays the foundation to fulfill the Naval Energy Resiliency, as outlined below:
   \begin{itemize}
     \item Address electrical power intermittency, integrate alternative energy sources into the grid, enhance grid security, develop energy storage solutions, promote local generation of zero-carbon fuels, and ensure inspection and health monitoring of critical energy infrastructure.
     \item Improve the ability to avoid, prepare for, minimize, adapt to, and recover from anticipated and unanticipated energy disruptions to ensure energy availability and reliability sufficient to provide for mission assurance, including mission essential operations related to readiness, and to execute or reestablish mission essential requirements.
     \item Enhance the ability of naval platforms around the world to accomplish their missions despite the actions by adversaries or other events to deny, disrupt, exploit, or destroy installation-based capabilities.
   \end{itemize}

\section*{Acknowledgment}
The information, data, or work presented herein was
funded in part by the U.S. Office of Naval Research
under award number N000142212239.

\section{Conclusion}
This paper presents a novel approach employing a deep graph transformer network for line-to-line fault diagnosis at the component level in the Naval MVDC 12kV system. By leveraging both spatial and temporal correlations within graph-based time-series data, the proposed network adeptly handles the complexities of localizing the faults, including successive multiple fault scenarios. The use of gated recurrent units further enhances the extraction of temporal features. Meanwhile, the multi-head attention mechanism of the graph transformer network improves the model’s ability to capture long-range dependencies and complex relationships by focusing on critical nodes across the entire graph rather than just local neighborhoods. The simulation of the shipboard system, encompassing all relevant components, helps to validate the effectiveness and reliability of the proposed approach. This innovation not only advances fault detection methodologies but also ensures reliability and accuracy in monitoring MVDC systems, marking a significant contribution to the field. Compared to bench-mark metrics, the proposed model achieves considerably better performance, even in noise scenarios.

\addtolength{\tabcolsep}{-3pt} 
\renewcommand{\arraystretch}{1.35} 

\begin{table}[htp]
\centering
\caption{Performance comparison with other schemes in the literature}
\begin{tabular}{c ccccc}
\toprule
\small{Method}  & \small{Accuracy} & \small{Case Study}\\
\midrule
\footnotesize{Multilevel Iterative} & \footnotesize{99.77\%} & \footnotesize{Binary fault diagnosis for} \\
\footnotesize{LightGBM \cite{liu2021fault}} & \footnotesize{ } & \footnotesize{shipboard MVDC system} \\
\hline
\footnotesize{Wavelet CNN \cite{senemmar2022non}} & \footnotesize{95\% - 97\%} & \footnotesize{Non-intrusive load monitor} \\
\footnotesize{ } & \footnotesize{ } & \footnotesize{classification in ship system}\\
\hline
\footnotesize{1-D Graph Attention} & \footnotesize{98.03\%} & \footnotesize{Fault management on} \\
\footnotesize{Network \cite{ngo2024deep}} & \footnotesize{ } & \footnotesize{power distribution systems}\\
\hline
\footnotesize{Generative Adversarial} & \footnotesize{ 91\% - 96\%} & \footnotesize{Fault localization on} \\
\footnotesize{Network \cite{liu2020deep}} & \footnotesize{ } & \footnotesize{the electric ship}\\
\hline
\footnotesize{Wavelets DNN \cite{li2014fault}} & \footnotesize{ 94.10\%} & \footnotesize{Fault detection and} \\
\footnotesize{} & \footnotesize{ } & \footnotesize{classification on SPS system}\\
\hline
\footnotesize{LSTM \cite{senemmar2021deep}} & \footnotesize{ 99.41\%} & \footnotesize{Fault diagnosis on 8-bus} \\
\footnotesize{} & \footnotesize{ } & \footnotesize{shipboard system}\\
\hline
\footnotesize{ResNet18 \cite{peng2024fault}} & \footnotesize{ 99.72\%} & \footnotesize{Fault classification on} \\
\footnotesize{} & \footnotesize{ } & \footnotesize{ MVAC shipboard system}\\
\hline
\footnotesize{Support Vector} & \footnotesize{ 95.63\%} & \footnotesize{Fault diagnosis on} \\
\footnotesize{Machine \cite{li2014fault}} & \footnotesize{ } & \footnotesize{Ship MVDC system}\\
\hline
\footnotesize{Wavelet GCN \cite{senemmar2024non}} & \footnotesize{ 94\% - 99\%} & \footnotesize{Fault detection in} \\
\footnotesize{} & \footnotesize{ } & \footnotesize{MVDC SPS}\\
\hline
\footnotesize{\textbf{Recurrent Graph}} & \footnotesize{ \textbf{99.69\%}} & \textbf{\footnotesize{Multiple Fault location on}} \\
\footnotesize{\textbf{Transformer Network}} & \footnotesize{ } & \textbf{\footnotesize{MVDC 12kV ship system}}\\
\bottomrule
\end{tabular}
\label{table:result4}
\end{table}
\bibliographystyle{IEEEtran}
\bibliography{bib.bib}

\end{document}